\documentclass[journal=aelccp,manuscript=letter]{achemso}

\usepackage{chemformula} % Formula subscripts using \ch{}
\usepackage[T1]{fontenc} % Use modern font encodings
\usepackage[disable=true]{todonotes}
\usepackage{hyperref}
\usepackage{physics}
\usepackage{xfrac}
\usepackage{amsmath}

\usepackage{enumitem,amssymb}
\newlist{todolist}{itemize}{2}
\setlist[todolist]{label=$\square$}
\usepackage{pifont}

\author{Maxwell D. Radin}
\email{hello@maxradin.com}
\author{Alexander Kunitsa}
\affiliation[Zapata AI]
{Zapata AI}

\title {Role of symmetry-forbidden transitions in resonant inelastic X-ray scattering from materials with trapped molecular \ch{O2}}

\abbreviations{}

\begin{document}

\begin{abstract}
Oxygen resonant inelastic X-ray scattering has become a prominent tool for unveiling the electronic structure of solids, especially redox mechanisms in Li-rich cathode materials.
A number of studies have observed a strong absorption feature at 531~eV associated with the anomalous capacity of Li-rich cathodes.
The most prominent emission feature arising from absorption at 531~eV occurs at 523~eV and has been attributed to the $^3\Pi_\text{g}$ final state of molecular \ch{O2}.
However, in many materials, this feature exhibits a secondary emission peak at a loss of 4.3~eV loss which is not seen in the spectrum of gas-phase oxygen.
This study revisits the spectra of these materials and investigates transitions that are forbidden for an isolated \ch{O2} molecule as a possible explanation for the secondary emission peak.
A group theoretic analysis and simulation of RIXS cross-sections suggest that this feature is consistent with molecular oxygen in a low-symmetry environment owing to final states descending from the $^3\Sigma_\text{u}^+$ and $^3\Delta_\text{u}$ states of isolated \ch{O2}.
Vibronic coupling and multi-molecule transitions are also discussed as possible mechanisms for enabling transitions to these final states even in a high-symmetry environment.
These results support the assignment of the 531~eV absorption peak in these materials exclusively to molecular oxygen and suggest that in materials exhibiting the secondary emission peak, these molecules may reside in low-symmetry environments.
\end{abstract}

Li-rich nickel-manganese-cobalt (NMC) oxides have attracted great interest as cathode materials because of the exceptionally high capacities they exhibit, greatly exceeding those expected from the conventional picture of transition-metal redox.\cite{Lu2002,Johnson2004}
However, their practical use is curtailed by a number of issues in electrochemical performance, including poor rate capability, large hysteresis, and significant voltage fade.\cite{Radin2019Manganese}
This has motivated many studies investigating the underlying redox mechanisms in these compounds as well as a much broader class of alkali transition-metal oxides that exhibit similar behavior.\cite{Radin2019Manganese,Yang2018,Wu2019,Lebens-Higgins2019,House2020,Vinckeviciute2021Two,Kitchaev2021delocalized,Zuba2021,Menon2023oxygen,Maitra2018,Dai2018,Hong2019}
An understanding of these mechanisms could potentially yield insights into how to improve the electrochemical performance of these compounds and realize their extraordinary capacities in practical applications.

A recurring pattern in these investigations is the appearance of a strong X-ray absorption peak at 531~eV during charging and associated emission at 523~eV in Li-rich NMCs.\cite{Yang2018,Wu2019,Lebens-Higgins2019,House2020,Zuba2021,Menon2025}
This behavior has also been observed in related electrochemical systems that exhibit a similar anomalous capacity, including Mn-based Na cathodes that exhibit \cite{Maitra2018,Dai2018} as well as iridium-based Li-rich cathodes \cite{Hong2019}.
This feature is broadly interpreted as an indication that some form of oxygen redox is responsible for the anomalous capacity of these materials, or alternatively, that oxygen is oxidized through beam damage or ex situ decomposition.\cite{Radin2019Manganese,Menon2025}
More recently, similar behavior has also been observed in conventional layered oxide cathodes at high states of charge, challenging the previous understanding of redox mechanisms in materials such as \ch{LiNiO2} and \ch{LiCoO2}.\cite{Menon2023oxygen,Juelsholt2024,Ogley2025}
A similar spectroscopic feature has also been observed in a range of as-prepared oxides, incliuding \ch{Al2O3}, \cite{Arhammar2011} \ch{Li2O},\cite{Zhuo2018} \ch{LiAlO2},\cite{Lebens-Higgins2019} and \ch{KMnO4}.\cite{Zuba2021}
In the latter two cases, the feature was demonstrated to be a consequence of beam damage.

A number of studies have explored what species may be responsible for this 531~eV feature and found strong evidence that it originates from molecular oxygen in many cases.
The absorption and emission energies are close to those of gas-phase \ch{O2} and high-resolution RIXS studies, both in Li-rich and conventional layered oxides as well as \ch{Al2O3}, have found a vibronic splitting of the elastic peak that agrees well with the vibrational modes of molecular \ch{O_2} \cite{Arhammar2011,House2020,Menon2023oxygen,Juelsholt2024,Ogley2025}.
Many questions however still remain.
For example, the origin of such molecular oxygen is not always clear, and has been suggested to arise in various contexts from reversible redox,\cite{House2020,Menon2023oxygen} beam damage,\cite{Radin2019Manganese,Lebens-Higgins2019,Zuba2021} and material synthesis.\cite{Zhuo2018} 

A potentially important clue into the origins of the 531~eV feature is that, despite the similarities described above, these compounds also exhibit some key differences in spectroscopic behavior.
First, in some cases, prolonged beam exposure induces the 531~eV feature, while in others, exposure removes it.\cite{Lebens-Higgins2019,Zuba2021}
Second, the 531~eV feature observed in solid-state systems occurs at an energy a few tenths of an eV above that of gaseous oxygen.\cite{Zhou2022}
Third, in some materials, a secondary emission peak is observed at 526.5~eV, whereas in gas-phase oxygen and some other materials it is absent or very weak.\cite{Zuba2021,Lebens-Higgins2019,Arhammar2011,Glans1996,Hennies2010,Menon2025,Zhou2022} 
Understanding the root cause of these differences could shed light on whether molecular oxygen is indeed wholly responsible for the 531~eV feature in these materials, and if so, what are the mechanisms by which it forms.

To shed light on the origin of this secondary emission peak, this study explores the possibility that it arises from transitions that are symmetry forbidden for isolated oxygen molecules.
A group theoretic analysis and simulations of RIXS cross-sections show that the 526.5~eV emission feature is consistent with molecular oxygen in a low-symmetry environment.
In particular, the feature can be explained by final states descending from the $^3\Sigma_\text{u}^+$ and $^3\Delta_\text{u}$ states, which, in high-symmetry environments, are forbidden by dipole selection rules.
These results support the notion that the 531~eV feature observed in various oxides is due to molecular oxygen and suggest that in materials exhibiting the secondary peak, such as charged Li-excess cathodes, the local environment of these molecules has a lower symmetry.
In contrast, materials that exhibit 531~eV absorption without the secondary peak, such as post-exposure \ch{KMnO4}, may hold oxygen in higher symmetry environments.
A testable prediction of this analysis is that the relative strength of features is expected to vary with the orientations of incoming/outgoing photons and their polarizations.

\begin{figure}
    \centering
    \includegraphics{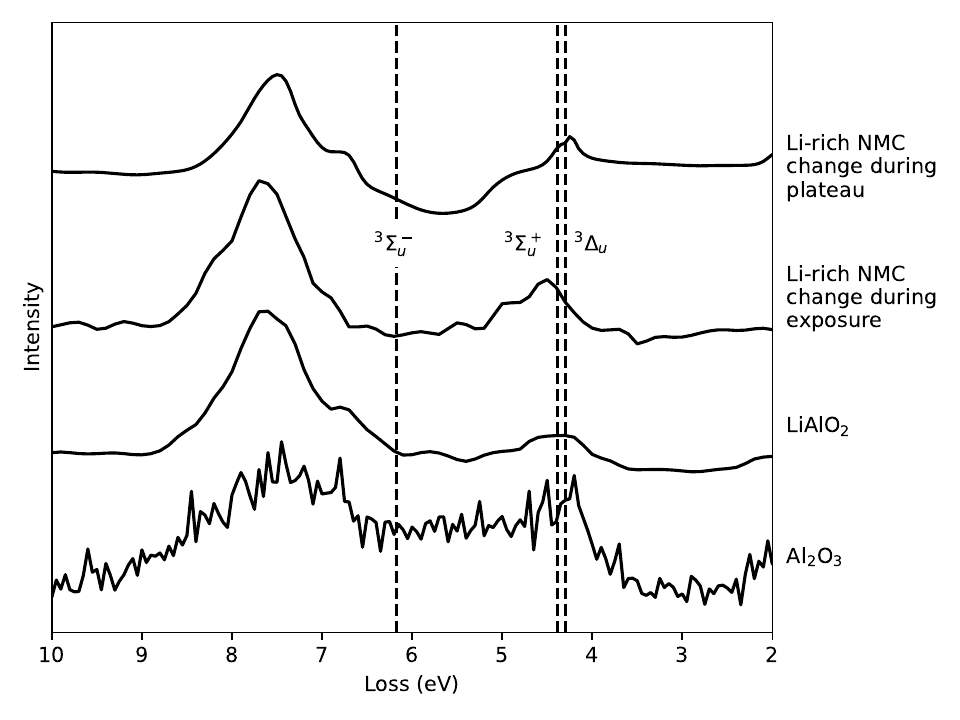}
    \caption{Comparison of experimental spectra of \ch{Al2O3}\cite{Arhammar2011} and \ch{LiAlO2}\cite{Lebens-Higgins2019} to the change in the spectrum of a charged Li-rich NMC cathode during the first-charge plateau.\cite{House2020}} The vertical lines indicate the minimum electronic energy $T_\text{e}$ determined from spectroscopic experiments\cite{NISTWebBook} for final states that are forbidden for isolated \ch{O2} molecules. Note that these lines do not include vibrational contributions and should be interpreted as an approximate upper bound on the loss energy.
    \label{fig:spectra}
    \todo{Check if NIST Te data is including zero-point energy}
\end{figure}

Fig. \ref{fig:spectra} illustrates the similarities in the emission spectra from materials that exhibit an absorption feature at 531~eV.
The bottom two spectra, obtained from \ch{Al2O3}\cite{Arhammar2011} and \ch{LiAlO2},\cite{Lebens-Higgins2019} show materials where both the primary 7.6~eV loss feature as well as the secondary 4.3~eV loss emission features are visible.
The top two spectra demonstrate the presence of a similar feature in charged cathodes when looking at spectrum differences.
The second line from the top shows the difference in the spectrum before and after a charged Li-rich NMC cathode is exposed to X-rays.\cite{Lebens-Higgins2019}
(Note that this spectrum corresponds to a component that was  removed by beam exposure, not a component that was created by beam exposure.\cite{Lebens-Higgins2019})
The top line shows the change in spectrum of a Li-rich NMC cathode during the first-charge plateau obtained by extracting spectra from Ref. \citenum{House2020}.
The striking similarities in spectra such as these suggest that the same species may be responsible. 

\begin{figure}
    \centering
    \includegraphics{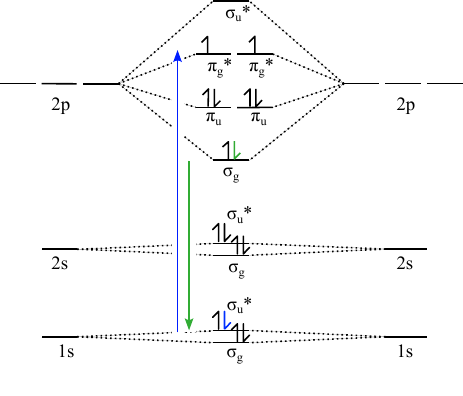}
    \caption{Transitions associated with the RIXS spectrum of an isolated oxygen molecule.}
    \label{fig:molecular-orbitals}
\end{figure}

The 531~eV absorption peak has been attributed in many cases to the excitation of a core 1s electron of molecular \ch{O2} into the lowest unoccupied state, a spin-down $\pi_g^*$ state derived from atomic 2p orbitals, resulting in a state of $^3\Pi_\text{u}$ symmetry.\cite{Arhammar2011,House2020,Menon2023oxygen,Frati2020}
The 7.6~eV loss feature is associate with relaxation of the $3\sigma_g$ spin-down electron into the core hole, yielding a final state with $^3\Pi_\text{g}$ symmetry.\cite{Hennies2010}
These transitions are illustrated in terms of molecular orbitals in Fig. \ref{fig:molecular-orbitals}.
Note that a close examination of Fig. \ref{fig:spectra} reveals a small feature at a loss of 6.8~eV in \ch{LiAlO2} and the change during the charging plateau of Li-rich NMC \cite{House2020}.
Although not discussed in these experiments, this feature also appears in gaseous oxygen and, like the 7.6~eV loss feature, is associated with the $^3\Pi_\text{g}$ final state.\cite{Hennies2010}
While the presence of the 6.8~eV loss feature confirms the presence of molecular oxygen in these systems, the origin of the emission peak at a 4.3~eV loss remains unclear.\cite{Menon2025}

Prior studies on other types of compounds motivate the possibility that this secondary emission feature could arise from some transition that is symmetry forbidden for isolated oxygen.
For example, there are many cases where transitions that ordinarily would be dipole forbidden can be enabled by interactions with additional degrees of freedom.
One example is the coupling to vibronic modes of the appropriate symmetry, as in the case of the O K-edge of \ch{CO2}.\cite{Maganas2014,Zhuo2020}
Another is multi-molecule processes, such as the $^3\Sigma_\text{g}^-$ to $^1\Delta_\text{g}$ transition responsible for the blue color of liquid oxygen.\cite{Krupenie1972,Tada2020,Biennier2000}
While this transition is dipole forbidden for a single molecule, it can occur as a one-photon two-molecule process.
Another type of mechanism, which will be the main focus of this study, is a local environment that breaks the symmetry of spectroscopically active species.
One example is the strong pre-edge peaks seen in the 1s X-ray spectra of transition-metal ions in environments lacking inversion symmetry (e.g., tetrahedral coordination).\cite{DeGroot2009,Zuba2021}
These are associated with excitation into the d manifold, a process that is forbidden by dipole selection rules in systems with inversion symmetry.
While the exact environment of trapped \ch{O2} remains clear, the geometries suggested by computational studies are generally low symmetry.\cite{Vinckeviciute2021Two,Juelsholt2024,Chen2016Lithium,McColl2022}
Altogether, the mechanisms described above motivate the consideration of transitions that are forbidden for an isolated oxygen as possible origins of the 4.3~eV loss feature.

Although in principle the oxygen molecule has many excited states, the most likely candidates for a final state are those that correspond to the de-excitation of a single electron to fill the core-hole.
(Multielectron photo-excitations are in principle possible but likely to yield much smaller cross-sections.)
Among these, final states that have a triplet configuration are likely to dominate over those with other spin states as only triplet final states are allowed under electric dipole transitions.
Considering the orbitals from which electrons could be de-excited, the $\sigma_u$ orbitals derived from atomic 2p orbitals can be immediately excluded as a possible explanation for the 4.3~eV loss feature given that it would yield emission around 503~eV \cite{Glans1996}, representing a loss of 18~eV.
The remaining possible final states correspond to the de-excitation of a $\pi_g$ electron.
These states form a four-dimensional subspace, corresponding to the two choices for which $\pi_u$ orbital the core electron is excited into and the two choices for which $\pi_g$ orbital an electron is de-excited from.
This subspace can be decomposed into the irreducible representations of $^3\Sigma_\text{u}^-$, $^3\Sigma_\text{u}^+$, and (doubly degenerate) $^3\Delta_\text{u}$ symmetry.
Note that transition from the $^3\Pi_u$ core-hole state to these four states are forbidden by dipole selection rules for an isolated oxygen molecule because the presence of inversion symmetry forbids transitions between states with the same parity.

Fig. \ref{fig:spectra} shows that the energies corresponding to $^3\Sigma_\text{u}^+$ and $^3\Delta_\text{u}$ final states, shown as vertical lines, agree well with the 4.3~eV loss observed experimentally.
These lines represent the minimum electronic energy $T_\text{e}$ for a given state as determined by spectroscopic measurements.\cite{NISTWebBook} 
(Note that vibrational effects could result in these states contribute to the loss at energies different from $T_\text{e}$, and in particular, contribute to losses at significantly larger energies due to vibrationally excited final state.) 
In contrast, no noticeable peaks appear at the energy corresponding to the $^3\Sigma_\text{u}^-$ final state.
This suggests that final states descending from the $^3\Sigma_\text{u}^+$ and/or $^3\Delta_\text{u}$ irreducible representations may be the source of the 4.3~eV loss feature.
A natural question then is what, if any, plausible local environments could yield significant cross-sections for the $^3\Sigma_\text{u}^+$ and/or $^3\Delta_\text{u}$ states without yielding a comparable cross-section for the $^3\Sigma_\text{u}^-$ state.

% There are a few possible explanations for why this final state would not appear in this spectrum despite the other forbidden final states, $^3\Sigma_\text{u}^+$ and $^3\Delta_\text{u}$, being visible.
% One is simply number of states: the $^3\Sigma_\text{u}^+$ state and doubly degenerate $^3\Delta_\text{u}$ state represent three possible final states corresponding to a 4.3~eV loss, which, all else being equal would like to a peak three times larger than the single $^3\Sigma_\text{u}^-$ state at a loss of ~6.2~eV.
% A second is differences in the transition dipole; the strength of each peak is dependent on the molecular wavefunctions of initial and final states.
% As an extreme case, in an environment with $D_{3\text{h}}$ symmetry, the dipole selection rules will permit transitions to final states corresponding to $^3\Sigma_\text{u}^+$ and $^3\Delta_\text{u}$ but not to the final state corresponding to $^3\Sigma_\text{u}^-$.

The RIXS cross section can be approximated from perturbation theory
\cite{Nanda2020} as
\begin{equation}
    \label{eqn:sigma}
    \sigma^\text{RIXS}\left(\theta\right) = 
    \frac{1}{15}\frac{\omega_e}{\omega_i}
    \left[
        \left(2-\frac{1}{2}\sin^2{\theta}\right) \nabla_g
        + \left(\frac{3}{4} \sin^2{\theta} -\frac{1}{2} \right) \left(\nabla_f + \nabla_h \right)
    \right]
\end{equation}
where $\theta$ is the angle between the polarization vector of the incoming photon and the propagation vector of the outgoing photon, $\omega_e$ and $\omega_i$ are the frequencies of the incoming and outgoing photons, and $\nabla_f$, $\nabla_g$, and $\nabla_h$ are given by \todo{Check if i and f are missing}
\begin{equation}
\begin{aligned}
\label{eqn:del-fgh}
&\nabla_f = \sum_{a,b} S_{aabb} \\
&\nabla_g = \sum_{a,b} S_{abab} \\
&\nabla_h = \sum_{a,b} S_{abba}.
\end{aligned}
\end{equation}
These represent the isotropic components of the RIXS transition strength tensor $S$, which is defined as\cite{Nanda2020}
\begin{equation}
    \label{eqn:s}
    S^{if}_{ab,cd} = \frac{1}{2}\left[
        M^{i \leftarrow f}_{ab} M^{f \leftarrow i}_{cd} +
        \left(M^{i \leftarrow f}_{cd}\right)^* \left(M^{f \leftarrow i}_{ab}\right)^*
    \right]
\end{equation}
where the left and right RIXS moments $M^{i \leftarrow f}$ and $ M^{f \leftarrow i}$, which within the equation-of-motion coupled cluster (EOM-CC) framework can be computed using Eqs. 10 and 11 of Ref. \citenum{Nanda2020}.
The calculations presented here employ the 6-311G(2df,2p) basis set and EOM-CC method for electronically excited states with single and double excitations (EOM-EE-CCSD)~\cite{krylov_equation--motion_2008}.

\begin{figure}
    \centering
    \includegraphics{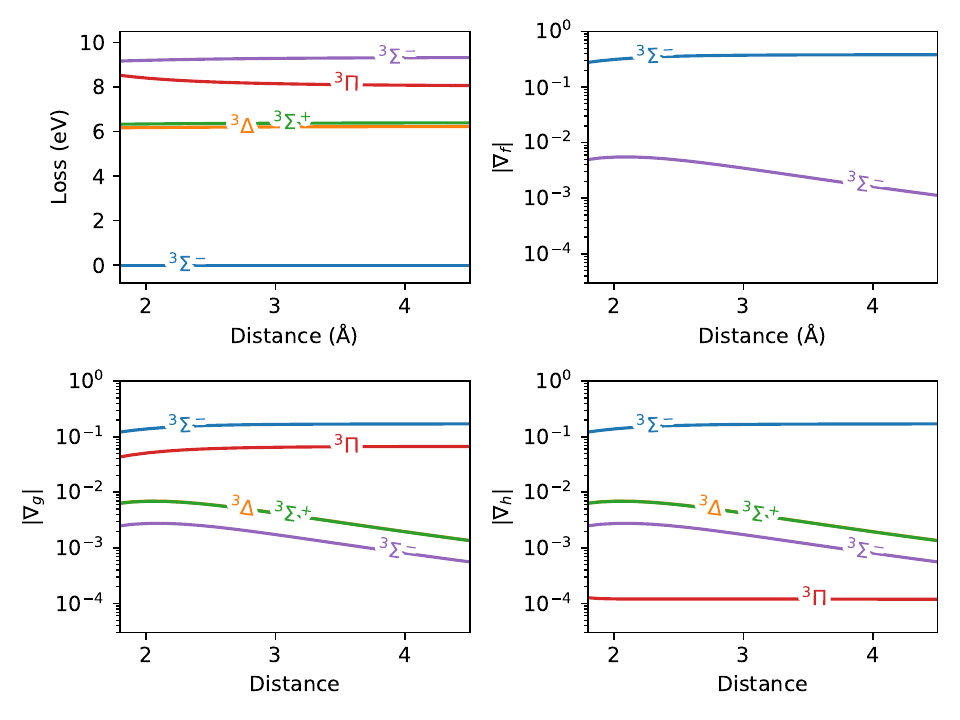}
    \caption{ Calculated energies and absolute RIXS components $|\nabla_f|$, $|\nabla_g|$, and $|\nabla_h|$ for an \ch{O2} molecule with a $+1$ point charge along the bond axis. Note that all terms are positive except for $\nabla_h$ for the $^3\Sigma^+$ and that the $^3\Delta$ and $^3\Sigma^+$ lines are overlapping for $|\nabla_g|$, and $|\nabla_h|$. Values that are zero to within machine precision are not shown. The x-axis represents the distance from the point charge to the closest oxygen atom.}
    \label{fig:rixs-components}
\end{figure}

The upper left panel of Fig. \ref{fig:rixs-components} shows the influence of a nearby $+1$ point charge along the bond axis on the energies of possible RIXS final states as calculated with Q-Chem \cite{qchem}.
This $\text{C}_{\infty \text{v}}$ geometry can be viewed as a simple model for the local environment of \ch{O2} trapped within a solid.
Based on the irreducible representations and excitation energies, the $^3\Delta$, $^3\Pi$, $^3\Sigma^+$ states shown in Fig. \ref{fig:rixs-components} can be identified as descendants of the $^3\Delta_\text{u}$, $^3\Pi_\text{g}$, and $^3\Sigma_\text{u}^+$ states of the $\text{D}_{\infty \text{h}}$ geometry.
The $^3\Sigma^-$ state shown in blue descends from the $^3\Sigma_\text{g}^-$ state and the $^3\Sigma^-$ state shown in purple from the $^3\Sigma_\text{u}^-$ state.

The top right panel of Fig. \ref{fig:rixs-components} shows the $|\nabla_f|$ RIXS strength obtained for this geometry with Q-Chem as a function of distance to the point charge.
Only the two $^3\Sigma^-$ states yield non-zero values for $\nabla_f$.
The $^3\Sigma^-$ state descending from the $^3\Sigma_\text{g}^-$ state (blue line) corresponds to the elastic peak and exhibits a large value of $|\nabla_f|$ that has little dependence on the distance to the point charge.
The $|\nabla_f|$ value for the $^3\Sigma^-$ state descending from the $^3\Sigma_\text{u}^-$ state (purple line), in contrast, is smaller and asymptotically approaches zero as the distance to the point charge increase.
This reflects the fact that the non-zero $\nabla_f$ value for this state is a result of symmetry breaking: as the distance to the point charge increases, the system more closely resembles the $\text{D}_{\infty \text{h}}$ point environment.

The bottom left and bottom right panels of Fig. \ref{fig:rixs-components} show the $|\nabla_g|$ and $|\nabla_h|$ RIXS strengths as a function of the distance to the point charge.
The $^3\Sigma^-$ state corresponding to the elastic peak and exhibits larger values of $|\nabla_g|$ and $|\nabla_h|$ than the other states at all distances.
The $^3\Pi$ state, corresponding to the 7.6~eV loss feature observed experimentally for gaseous oxygen, exhibits the next largest values of $|\nabla_g|$ and $|\nabla_h|$.
The other states considered, however, all exhibit non-zero values of $\nabla_g$ and $\nabla_h$ as a result of symmetry breaking, reflected in an asymptotic approach to zero in the limit of large distance.

Many aspects of these results can be rationalized from symmetry considerations.
For example, in the case of the  $^3\Sigma^+$ final state, $\nabla_f$ vanishes because the diagonal entries of the RIXS moments, $M^{i \leftarrow f}_{aa}$ and $M^{f \leftarrow i}_{aa}$, are zero.
The $zz$ entries are zero because the $z$ dipole operator does not couple states of $^3\Pi$ symmetry to $^3\Sigma^+$.
To see why the $xx$ and $yy$ entries are also zero, consider that the $^3\Pi$ manifold is spanned by two states, one which is symmetric under a $\sigma_x$ reflection and one which is anti-symmetric under $\sigma_x$.
Because the $x$ dipole operator cannot couple states that are both symmetric or both anti-symmetric under a $\sigma_x$ reflection, neither of these two $^3\Pi$ states will be coupled by $x$ to both the initial and final states: $^3\Sigma^-$ is symmetric under $\sigma_x$ whereas $^3\Sigma^+$ is anti-symmetric.
Therefore the $xx$ entries must be zero, and by symmetry, the $yy$ entries must also vanish.
% For $^3\Delta$, dipole selection rules similarly require $zz$ to be zero.
% The $xx$ and $yy$ entries are both non-zero but do not contribute to $\nabla_f$ because they differ in sign.
% A detailed analysis of this cancellation of terms is beyond the scope of this manuscript, but we note that it can be interpreted as a consequence of wavefunction antisymmetrization and the fact the $xx$ and $yy$ entries correspond to de-exciting electrons from different $\pi_u$ molecular orbitals: one derived from atomic $p_x$ orbitals and the other from atomic $p_y$ orbitals.

% The RIXS moment $M^{i \leftarrow f}_{ab}$ transforms as $\Gamma_i \otimes \Gamma_f \otimes \Gamma^\text{dipole}_a \otimes \Gamma^\text{dipole}_b$ where $\Gamma_i$ and $\Gamma_f$ are the representations of the initial and final states, and $\Gamma^\text{dipole}_a$ and $\Gamma^\text{dipole}_b$ are the representations of the Cartesian coordinate variables corresponding to the $a$ and $b$ axes. \cite{Long2002}
% Note that M is a rank 2 Cartesian tensor, and only the Sigma^+ components contribute to del f.

\begin{figure}
    \centering
    \includegraphics{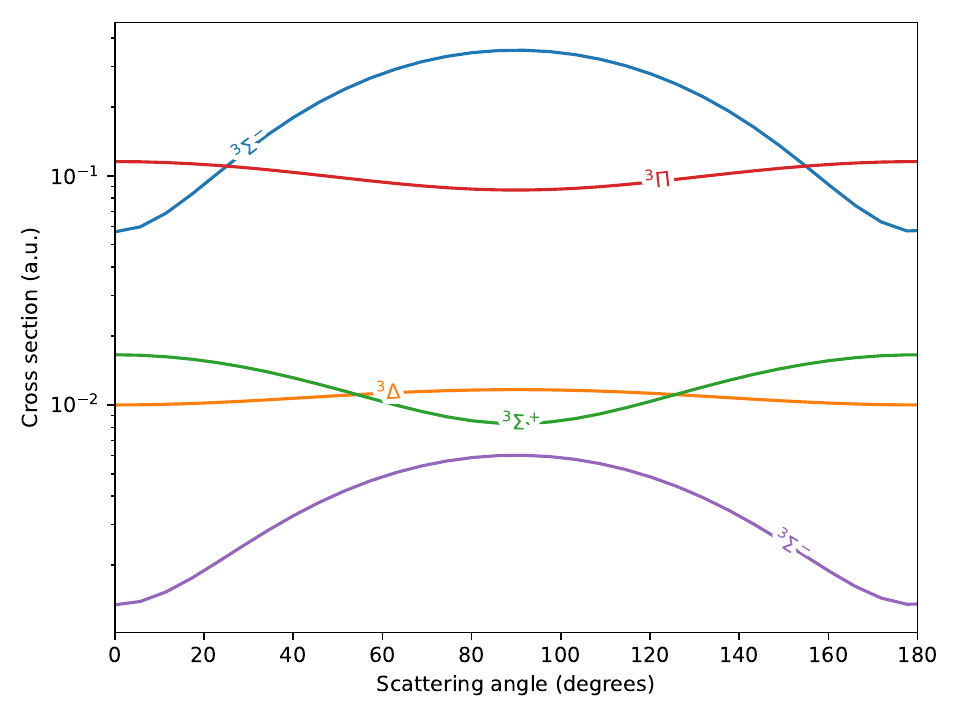}
    \caption{RIXS cross section for a \ch{O2} molecule with a $+1$ point charge along the bond axis 2.3~{\AA} from the nearest oxygen atom as a function of the angle between the polarization vector of the incoming photon and the propagation vector of the outgoing photon.}
    \label{fig:rixs-cross-section-vs-angle}
\end{figure}

Fig. \ref{fig:rixs-cross-section-vs-angle} shows the scattering cross vs. the angle $\theta$ between the polarization vector of the incoming photon and the propagation vector of the outgoing photon calculated from Eq. \ref{eqn:sigma} for the $\text{C}_{\infty \text{v}}$ geometry with the point charge at a distance of 2.3~\AA.
This distance corresponds to the Li-O distance in the geometry obtained for trapped \ch{O2} in Ref. \citenum{Vinckeviciute2021Two}.
For all values of $\theta$, the $^3\Pi$ and elastic $^3\Sigma^-$ states dominate, while the $^3\Sigma^+$ and $^3\Delta$ states yield cross sections about one order of magnitude smaller than that of the $^3\Pi$ state.
The $^3\Sigma^-$ state descending from $^3\Sigma^-_\text{u}$ has a cross-section that is significantly smaller than that of $^3\Sigma^+$ and $^3\Delta$.

The results above indicate that the experimental spectra shown in Fig. \ref{fig:spectra} are qualitatively consistent with molecular oxygen in a low-symmetry environment.
The experimental spectra show significant strength at the energies corresponding to the minimum electronic energy of the  $^3\Sigma^+_\text{g}$ and $^3\Delta_\text{g}$ and little weight at the energy corresponding to $^3\Sigma^-_\text{u}$.
This is consistent with the relative scattering strengths shown in Fig. \ref{fig:rixs-cross-section-vs-angle} and the fact that Fig. \ref{fig:rixs-components}a shows that the presence of the point charge has little effect on the energies of the $^3\Pi$, $^3\Delta$, and $^3\Sigma^+$ states.
Note that although the exact relationship between the incoming and outgoing polarization and propagation vectors is unclear in many experimental papers, Fig. \ref{fig:rixs-cross-section-vs-angle} indicates that the relative cross-sections of the $^3\Pi$, $^3\Sigma^+$ and $^3\Delta$ states do not vary significantly with $\theta$.

\begin{figure}
    \centering
    \includegraphics{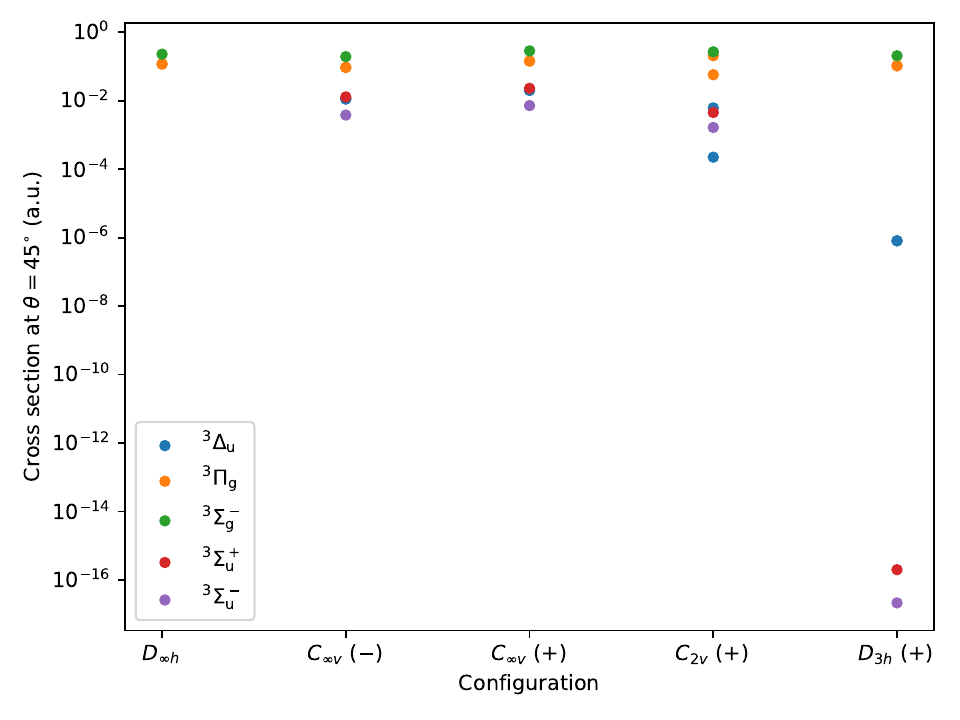}
    \caption{
        Comparison of calculated RIXS cross sections for several configurations.
        The labels on the horizontal axis indicate the point group of the configuration and the charge (positive or negative) on the point charges.
        Cross sections are shown for a distance of 2.1~{\AA} between the point charges and nearest oxygen atom and a $45^{\circ}$ angle between the polarization vector of the incoming photon and the propagation vector of the outgoing photon.
        States are labeled according to the irreducible representation of the $D_{\infty h}$ point group from which they descend.
    }
    \label{fig:geometry-comparison}
\end{figure}

To illustrate the sensitivity to local environment, Fig. \ref{fig:geometry-comparison} compares the calculated RIXS cross sections for several different geometries involving one or more point charges of charge $\pm 1$ to the cross section for an isolated \ch{O2} molecule ($\text{D}_{\infty \text{h}}$ symmetry).
In all cases the distance from the point charge to the nearest oxygen atom is fixed to 2.1~{\AA}.
The linear $\text{C}_{\infty \text{v}}$ geometry yields a significant contribution to the cross section from states that descend from the $^3\Sigma_\text{u}^+$ and $^3\Delta_\text{u}$ states, regardless of whether the coordinating point charge is positive or negative.
A $\text{C}_{2\text{v}}$ geometry consisting of a single positive charge that is equidistant from the two oxygen atoms yields cross sections for states descending from $^3\Sigma_\text{u}^+$, $^3\Sigma_\text{u}^-$, and $^3\Delta_\text{u}$ that are still significant but notably smaller than those in the $\text{C}_{\infty \text{v}}$ geometry.
In contrast, a $\text{D}_{3\text{h}}$ geometry consisting of three positive point charges lying in the horizontal plane equidistant from the two oxygen atoms (similar to oxygen dimer environments in compounds such as \ch{Li2O2}\cite{Foppl1957,Cota2005} and \ch{Na2O2}\cite{Tallman1957}) yields much smaller contributions to the cross-section from states descending from $^3\Sigma_\text{u}^+$, $^3\Sigma_\text{u}^-$, and $^3\Delta_\text{u}$.

The modest cross section of these final states in $\text{D}_{3\text{h}}$ configuration can be rationalized from group theoretic considerations.
Despite the lack of inversion symmetry in this point group, final states descending from $^3\Sigma_\text{u}^+$ and $^3\Sigma_\text{u}^-$ irreps are nevertheless forbidden.
To see this, first consider that the intermediate $^3\Pi_u$ state descends to the $\text{E}'$ irrep and the $^3\Sigma_\text{u}^+$ and $^3\Sigma_\text{u}^-$ states to $\text{A}_1''$ and $\text{A}_2''$.
Consulting the multiplication table for $\text{D}_{3\text{h}}$ shows that transitions from $\text{E}'$ to $\text{A}_1''$ and $\text{A}_2''$ are forbidden, i.e. the product of $\text{E}'$ with either of these does not contain the irreps corresponding to the $x$, $y$, or $z$ dipole operators.
One subtlety is that there is in principle another possible intermediate state at an energy similar to that of the $\text{E}'$ intermediate state.
This state corresponds to the excitation of an electron from an orbital descending from $1\sigma_u$ rather than $1\sigma_g$, and has  $\text{E}''$ symmetry in $\text{D}_{3\text{h}}$.
Because transitions from the ground state, which has $\text{A}_2'$ symmetry, to $\text{E}''$ are forbidden, this intermediate state will not contribute to the cross section.

Although final states descending from $^3\Delta_\text{u}$ are dipole allowed in $\text{D}_{3\text{h}}$, the low cross-section compared to the other geometries considered can be attributed to the fact that this final state is still forbidden within the single-particle picture.
The $\pi_u$ orbital descends to the $\text{E}'$ irrep while the $\sigma_u^*$ orbital descends to $\text{A}_2''$.
Because their direct product $\text{E}' \otimes \text{A}_2'' = \text{E}''$ does not contain the irreps corresponding to $x$, $y$, or $z$ dipole operators, a single-particle transition between these orbitals is dipole forbidden.
The presence of a non-zero cross section for the $^3\Delta_\text{u}$ final state can be attributed to contributions from other electron configurations.
For example, both $^3\Delta_\text{u}$ and $^3\Pi_\text{g}$ descend to the $\text{E}''$ irrep in $D_{3\text{h}}$ and therefore $^3\Delta_\text{u}$ can contain some weight from electron configurations such as the
$
\left(1\sigma_g\right)^2
\left(1\sigma_u^*\right)^2
\left(2\sigma_g\right)^2
\left(2\sigma_u^*\right)^2
\left(3\sigma_g\right)^1
\left(1\pi_u\right)^4
\left(1\pi_g^*\right)^3
$ final state shown in Fig. \ref{fig:molecular-orbitals}.

The analysis above bears some analogy to the O K-edge RIXS spectrum of \ch{Li2O2}, which also exhibits secondary features related to symmetry breaking. 
As shown by simulations and experiments,\cite{Zhuo2018} \ch{Li2O2} exhibits excitonic features associated with the excitation of an electron into the unoccupied $3\sigma^*_u$ orbital followed by the relaxation of $3\sigma_g$, $1\pi_u$, and $1\pi_g^*$ electrons into the core hole.
Although the local peroxide environment (point group $\text{D}_{3\text{h}}$) lacks inversion symmetry, the final states corresponding to de-excitation from $3\sigma_g$ and $1\pi_g^*$ orbitals (on the same peroxide dimer as the core hole) are nevertheless forbidden.
To see this, consider that de-excitation from the $3\sigma_g$ orbital, which corresponds to the
\begin{equation}
\left(1\sigma_g\right)^2
\left(1\sigma_u^*\right)^2
\left(2\sigma_g\right)^2
\left(2\sigma_u^*\right)^2
\left(3\sigma_g\right)^1
\left(1\pi_u\right)^4
\left(1\pi_g^*\right)^4
\left(3\sigma_u^*\right)^1
\end{equation}
electron configuration, descend from $\text{A}_{1u}$ to $\text{A}_2''$ when going from point group $\text{D}_{\infty \text{h}}$ to $\text{D}_{3\text{h}}$.
The same is true for the final state corresponding to de-excitation from the $1\pi_g^*$ orbital,
\begin{equation}
\left(1\sigma_g\right)^1
\left(1\sigma_u^*\right)^2
\left(2\sigma_g\right)^2
\left(2\sigma_u^*\right)^2
\left(3\sigma_g\right)^2
\left(1\pi_u\right)^4
\left(1\pi_g^*\right)^4
\left(3\sigma_u^*\right)^1.
\end{equation}
The final state corresponding to de-excitation from $1\pi_g^*$, 
\begin{equation}
\left(1\sigma_g\right)^2
\left(1\sigma_u^*\right)^2
\left(2\sigma_g\right)^2
\left(2\sigma_u^*\right)^2
\left(3\sigma_g\right)^2
\left(1\pi_u\right)^4
\left(1\pi_g^*\right)^3
\left(3\sigma_u^*\right)^1,
\end{equation}
instead descends from $\text{E}_{1u}$ to $\text{E}'$.
The fact that the direct products $\text{A}_2'' \otimes \text{A}_2'' = \text{A}_1'$ and $\text{A}_2'' \otimes \text{E}' = \text{E}''$ do not contain the irreps corresponding to the $x$, $y$, or $z$ dipole operators indicates that these final states are forbidden.
(Note that there is another potential intermediate state corresponding to the excitation of a $1\sigma_u$ electron, but in $\text{D}_{3\text{h}}$, this state does not couple to the initial state because both descend to $\text{A}_1'$.)

The fact that final states arising from the de-excitation of $3\sigma_g$ and $1\pi_g^*$ electrons appear in the experimental and theoretical RIXS spectra for \ch{Li2O2} despite being dipole forbidden for a single peroxide dimer in a $\text{D}_{3\text{h}}$ environment suggests that these features arise from the charge transfer between dimers.
For example, one can take linear combinations of $1\pi_g^*$ orbitals on a ring of six peroxide dimers to obtain an orbital that has $\text{A}_2''$ symmetry with respect to the dimer at the center of the ring.
De-excitation from such an orbital into a core-hole on the central dimer corresponds to a final state of $\text{A}_1'$ symmetry, which is coupled to the  $\text{A}_2''$ intermediate state via the $z$ dipole operator.
Although this hypothesis could be tested by, for example, examining the excitonic wavefunctions generated by simulations such as those in Ref. \citenum{Zhuo2018}, such modeling in beyond the scope of this work.

In summary, calculations suggest that the 4.3~eV loss feature observed in the RIXS spectrum of many oxides, including cathodes at high states of charge, are consistent with molecular oxygen in an environment that supports transitions that otherwise would be dipole forbidden.
A close examination of the spectrum of gas-phase oxygen supports this: a small but non-zero loss observed around 3-4~eV.\cite{Hennies2010,Zhou2022}
These results confirm that molecular oxygen is responsible for the 531~eV absorption peak observed in these systems and provide some insight into the local environment.
In particular, the simulations and group theoretic analysis show that the observed secondary peak is consistent with environments lacking inversion symmetry, and especially those breaking the horizontal reflection symmetry.
However it is difficult to rule out other possible mechanisms that could enable the $^3\Sigma_\text{u}^+$ and $^3\Delta_\text{u}$ final states.
This could include an environment that provides vibronic modes of the appropriate symmetry or environments that enable charge transfer between nearby \ch{O2} dimers as appears to occur in \ch{Li2O2}.
Calculations suggest that additional confirmation could be obtained by varying the experimental geometry, i.e. the relative orientations of incoming and outgoing polarization and propagation vectors.

\begin{acknowledgement}

The authors thank L. Piper and A. Menon for their sharing their expertise on the design of RIXS experiments.

\end{acknowledgement}

\bibliography{main}

\end{document}